\documentstyle[12pt,epsf]{article}
\newcommand{\bq}{\begin{equation}}
\newcommand{\eq}{\end{equation}}
\newcommand{\bqn}{\begin{eqnarray}}
\newcommand{\eqn}{\end{eqnarray}}
\newcommand{\nb}{\nonumber}
\newcommand{\lb}{\label} 

\begin{document} 
\baselineskip 0.9cm

\title{Gravitational collapse of massless
scalar field and radiation fluid }
\author{Anzhong Wang\thanks{Internet: wang@symbcomp.uerj.br;
wang@on.br} \\
\small Departamento de F\' {\i}sica Te\' orica,
Universidade do Estado do Rio de Janeiro, \\
\small Rua S\~ ao Francisco Xavier 524, Maracan\~ a,
20550-013 Rio de Janeiro~--~RJ, Brazil\\
\small and\\
\small Observat\'orio Nacional~--~CNPq, \\ 
\small Rua General Jos\'e Cristino 77, S\~ao Crist\'ov\~ao,
20921-400 Rio de Janeiro~--~RJ, Brazil\\
     \\
J.F. Villas da Rocha\thanks{Internet: roch@on.br}$\;$, $\;$ 
N.O. Santos\thanks{Internet: nos@on.br} \\
\small Departmento de
Astrof\'{\i}sica, Observat\'orio Nacional~--~CNPq, \\ 
\small Rua General Jos\'e Cristino 77, S\~ao Crist\'ov\~ao, 20921-400 
Rio de Janeiro~--~RJ, Brazil}
\date{}
\maketitle
\input epsf

\newpage
 
\begin{abstract}

\baselineskip 0.9cm

Several classes of conformally-flat and spherically symmetric exact
solutions to the Einstein field equations coupled with either a massless
scalar field or a radiation fluid are given, and their main properties
are studied. It is found that some represent the formation of black
holes due to the gravitational collapse of the matter fields. When the
spacetimes have continuous self-similarity (CSS), the masses of black
holes take a scaling form $M_{BH} \propto (P - P^{*})^{\gamma}$, where
$\gamma = 0.5$ for massless scalar field and $\gamma = 1$ for radiation
fluid.  The reasons for the difference between the values of $\gamma$
obtained here and those obtained previously are discussed. When the
spacetimes have neither CSS nor DSS (Discrete self-similarity), the
masses of black holes {\em always} turn on with finite non-zero values.

\end{abstract}

\vspace{.4cm}


\newpage

\baselineskip 0.9cm

\section*{I. INTRODUCTION}

Gravitational collapse is one of the fundamental problems in General
Relativity (GR).  The collapse generally has three kinds of possible
final states.  The first is simply the halt of  the processes in a
self-sustained object or the dispersion of a matter or gravitational
field.  The second is the formation of black holes with outgoing
gravitational radiation and matter, while  the third is the formation of
naked singularities. For the last case, however, the cosmic censorship
hypothesis \cite{Penrose1969}  declares that  these naked singularities
do not occur in Nature.  The study of gravitational collapse  has been
mainly guided by these three possibilities.

However, due to the mathematical complexity of the Einstein field
equations, we are frequently forced to impose some symmetries on the
concerned systems in order to make the problem tractable.  Spacetimes
with spherical symmetry are one of the cases. In particular, the
gravitational collapse of a minimally coupled massless scalar field in
such spacetimes has been studied both analytically \cite{Chris1986} and
numerically \cite{GP1987}, and some fundamental theorems have been
established.  Quite recently this problem has further attracted
attention, due to Choptuik's discovery of critical phenomena that were
hitherto unknown \cite{Ch1993}.  By using a very sophisticated method,
Choptuik showed numerically the following intriguing features: Let the
initial distribution of the massless scalar field be parameterized
smoothly by a parameter $P$ that characterizes the strength of the
initial conditions, such that the collapse of the scalar field with the
initial data $P > P^{*}$  forms a black hole, while the one with  $P <
P^{*}$ does not. Then, it was found that:  a) the critical solution with
$P = P^{*}$ is {\em universal} in the sense that in all the
one-parameter families of the solutions considered it approaches an
identical spacetime; b) the critical solution is periodic in the
logarithm of spacetime scale, with a period of $\triangle \approx 3.44$.
This is usually referred to as ``echoing" or discrete self-similarity
(DSS); c) near the critical solution (but with $P > P^{*}$),
the black hole mass is given by 
$$
M_{BH} = K (P - P^{*})^{\gamma},
$$
where $K$ is a family-dependent constant, but $\gamma$ is an apparently
{\em universal} scaling exponent, which has been numerically
determined as $\gamma \approx 0.37$.  These phenomena were soon  found
also in the collapse of axisymmetric gravitational waves \cite{AE1994}
and radiation fluid \cite{EC1994}.  Therefore, it
seems that the phenomena are not due to the particular choice of the
matter fields, but rather are generic features of GR. Further numerical
evidence to support this conclusion can be found in
\cite{EH1995}.

Parallel to the above numerical investigations, there have been
analytical efforts in understanding the physics behind these phenomena
\cite{Brady1994,Koike1995,Gundlach1995}. While the universality of the
critical solution and its self-similarity (echoing) have been found in
most cases, the universality of the  exponent $\gamma$ does not.  In
particular, Maison \cite{Maison1995} showed that $\gamma$ is
matter-dependent. For the collapse of the perfect fluid with the
equation of state $p = k \rho$, it strongly depends on $k$, where $p$
and $\rho$ are respectively the pressure and energy density of the
fluid, and $k$ is a constant.  The same conclusion was also reached both
analytically \cite{KO1995} and numerically \cite{EHH1995}.  Thus, one
might expect that $\gamma$ is universal only within a particular family
of matter fields.

However, even in this sense $\gamma$ is still not universal. Lately,
Oliveira and one of the present authors \cite{WO1996} constructed an
analytic model that represents the collapse of massless scalar wave
packets by using the so-called ``cut-paste" method to the model studied
in \cite{Brady1994}, and found that $\gamma = 0.5$ for spacetimes with
continuous self-similarity (CSS). This is different from the value
$\gamma \approx 0.374$ for spacetimes with DSS \cite{Ch1993}. Thus, the
exponent $\gamma$ depends not only on matter fields but also on
self-similarities, continuous or discrete \footnote{Note that the
original Roberts solutions \cite{Roberts1989} are not regular at the
center $R = 0$. However, as shown in \cite{Brady1994}, for the
subcritical solutions the hypersurface $R = 0$ is always time-like and
with negative mass. Since the past and future self-similarity horizons
carry flat-space null data, one can replace the negative mass part of
the spacetime with flat one in both the past and future light cones of
the singularity, so that the pieced spacetime has a regular center
\cite{Gundlach1995}. Of course, this will give up the analytic
condition, by which, together with the condition of a regular center, it
was shown that the critical solution found in \cite{AE1994} is unique.
In sequel, the exponent $\gamma$ is uniquely determined
\cite{Koike1995,Gundlach1995}.}.  A natural question now is: What will
happen when the collapse has neither CSS nor DSS?

In this paper we shall first present several classes of exact solutions
to Einstein's field equations coupled either with a massless scalar
field or with a radiation fluid, and then study their physical
properties. To derive these solutions, we assume that the spacetimes are
spherically symmetric and conformally flat. One might argue that
spacetimes with conformal flatness are not very realistic, and the total
mass of the spacetime is usually infinite. To overcome this shortage,
one may match the spacetimes to an asymptotically flat exterior by the
 so-called ``cut-paste" method \cite{WO1996}. In the present paper, we
shall briefly discuss this possibility, and leave the details to be
reported somewhere else.  The rest of the paper is organized as follows:
In section $II$ several classes of exact solutions to the Einstein field
equations coupled with either a massless scalar field or a radiation
fluid are derived, while in section $III$ their physical interpretations
are studied. The paper is closed with section $IV$ where our main
conclusions are presented.

\section*{ II. ANALYTIC SOLUTIONS OF MASSLESS SCALAR 
FIELD AND RADIATION FLUID}
 
\renewcommand{\theequation}{2.\arabic{equation}}
\setcounter{equation}{0}

The general spherically symmetric spacetime is described by the
metric \cite{LL1975}
$$
ds^{2} = G(t, r) dt^{2} + 2 H(t, r)dt dr - J(t, r)dr^{2}
- K(t, r) d\Omega^{2},
$$
where $d\Omega^{2} = d\theta^{2} + \sin^{2}\theta d\varphi^{2}$, and
$\{x^{\mu}\}\equiv \{t, r, \theta, \varphi\}\; (\mu = 0, 1, 2, 3)$ are
the usual spherical coordinates. Due to the arbitrariness in the choice
of coordinates, it is subject to the following coordinate
transformation, $t = T(t', r'), \;r = R(t', r').$ Making use of this
freedom,  we can set $ H(t, r) = 0$. if we further consider the
shear-free case \cite{BOS1989}, we have  $\; K(t, r) = r^{2} J(t, r)$.
 Then, the metric takes the form
$$
ds^{2} = G(t, r) dt^{2} -  K(t, r)\left(dr^{2}
+ r^{2} d\Omega^{2}\right).
$$
Note that the above metric is still subject to the transformation
$\bar{t} = f(t)$, where $f(t)$ is an arbitrary function. Later on, we
shall use this freedom to further simplify the metric.  With the above
form of metric, one can show that the conformal-flatness condition
$C_{\mu\nu\lambda\sigma} = 0$, where $C_{\mu\nu\lambda\sigma}$ denotes
the Weyl tensor, now reads
$$
C,_{rr} - \left(\frac{C}{r}\right),_{r} - \frac{C}{r^{2}} = 0,
$$
where $C \equiv \sqrt{G(t, r)/K(t, r)}$, and $( ),_{x} \equiv \partial(
)/\partial x$. The above equation has the general solution
$$
C(t, r) = f_{1}(t) + f_{2}(t) r^{2},
$$
where $f_{1}$ and $f_{2}$ are two arbitrary functions of $t$. Thus, 
 there are three possibilities: 
$$
i)\; f_{1}(t)\not= 0, f_{2}(t) =
0, \;\;\; ii)\; f_{1}(t)= 0, f_{2}(t) \not= 0,\;\;\;
iii)\; f_{1}(t)\not=
0, f_{2}(t) \not= 0.
$$
In case i), by introducing a new coordinate $\bar{t} \equiv
\int{f_{1}(t) dt}$ we can bring the metric to a form that is
conformally flat to the Minkowski metric.   Thus, without loss of
generality, in this case we can set $f_{1}(t) = 1$. By a similar
argument, we can set $f_{2}(t) = 1$ in cases ii) and iii). Once this is
done, cases i) and ii) are not independent. In fact,  by a coordinate
transform $r = 1/\bar{r}$, the metric of case ii) will reduce to that of
case i). Therefore,  the general metric for spherically symmetric
spacetimes with conformal flatness takes the form
\bq
\lb{eq2.1}
ds^{2} = G(t, r) \left[ dt^{2} - h^{2}(t, r)\;\left(dr^{2} 
+ r^{2}d\Omega^{2}\right)\right],
\eq
where
\bq
\lb{eq2.2}
h(t, r) = \frac{1}{C(t, r)} = \left\{
\begin{array}{c}
1, \\
\frac{1}{f_{1}(t) + r^{2}},
\end{array}
\right.
\eq
with $f_{1}(t) \not= 0$. In the following, we shall refer solutions
with $C(t, r) = 1$ to as Type $A$ solutions, and solutions with $C(t, r) =
f_{1}(t) + r^{2}$ to as Type $B$ solutions. 

The concept of CSS (or homotheticity) is defined in a relativistic
context as the existence of a conformal Killing vector field
$\xi^{\mu}$, satisfying \cite{CT1971}
$$
\xi_{\mu;\nu} + \xi_{\nu;\mu} = 2 g_{\mu\nu}.
$$
Because of the spherical symmetry, we can write $\xi^{\mu}$ as
$\xi^{\mu} = \xi^{0}\delta^{\mu}_{t} + \xi^{1}\delta^{\mu}_{r}$,
where $\xi^{0}$ and $\xi^{1}$ are functions of $t$ and $r$. Substituting
this expression into the above equations, we find
\bqn
\lb{eq2.3}
\xi^{1} R,_{r} + \xi^{0} R,_{t} &=& R, \nb\\
\xi^{1}\nu,_{r} + \xi^{0}\nu,_{t} + \xi^{1},_{r} &=& 1, \nb\\
\xi^{1}\lambda,_{r} + \xi^{0}\lambda,_{t} + \xi^{0},_{t} &=& 1,\nb\\
e^{2\nu} \xi^{1},_{t} - e^{2\lambda}\xi^{0},_{r} &=& 0, 
\eqn
where $\lambda \equiv \frac{1}{2}\ln{G},\; \nu \equiv
\frac{1}{2}\ln({h^{2}G})$, and $R \equiv rhG^{1/2}$.

The concept of DSS is defined as follows \cite{Gundlach1995}: If there
exists a diffeomorphism $\phi$ and a real constant $\triangle$ such that,
for any integer $n$,
$$
\left(\phi_{*}\right)^{n} g_{ab} = e^{2n\triangle} g_{ab},
$$
then the corresponding spacetime is said to have DSS, where $\phi_{*}$
denotes the pullback of $\phi$. For the metric (\ref{eq2.1}) it can be
shown that the diffeomorphism implies that 
\bq
\lb{eq2.4}
G(t, r) = G(e^{n \Delta} t, e^{n \Delta} r),\;\;\;\;
h(t, r) = h(e^{n \Delta} t, e^{n \Delta} r).
\eq

To see the connection between CSS and DSS, one may define a vector field
$\xi \equiv \partial /\partial \sigma$, where $\sigma$ is one of the
four coordinates of the spacetime such that if a point $p$ has the
coordinate $(\sigma, x^{a})$, its image $\phi(p)$ has the coordinate
$(\sigma + \triangle, x^{a})$. The discrete diffeomorphism $\phi$ is then
realized as the Lie dragging along $\xi$ by a distance $\triangle$.
Clearly, CSS corresponds to DSS for infinitesimally small $\triangle$. In
this sense, CSS can be considered as a degenerate case of DSS.  For the
details, we would like to refer readers to \cite{Gundlach1995}.

The functions $G(t, r)$ and $f_{1}(t)$ are determined by the Einstein
field equations $R_{\mu\nu} - g_{\mu\nu}R/2 = - 8\pi T_{\mu\nu}.$ Note
that in this paper we choose units such that $G = 1 = c$, where $G$
is the gravitational constant, and $c$ the speed of light. For
the metric (\ref{eq2.1}), the non-vanishing components of the Ricci
tensor are given by
\bqn
\lb{eq2.5}
R_{00} &=& \frac{3 }{2 h}
 \left(  \frac{ h G,_{t} }{G}\right),_{t}  + \frac{ 3 h,_{tt}}{h} 
- 
\frac{(hr^2 G,_{r}),_{r}}{ 2r^{2} h^{3} G  }, \\
\lb{eq2.6}
R_{01} &=&  \frac{h}{G}
 \left( \frac{G,_{r} }{h}\right),_{t}
- \frac{3G,_{t}G,_{r}}{2G^{2}}  
+ \left(\frac{2h,_{r}}{h} \right),_{t}, \\
\lb{eq2.7}
R_{11} &=&  \left( \frac{3G,_{r}}{2 G} \right),_{r} + 
\frac{ hG,_{r}}{2r^2 G} \left(\frac{ r^2 }{h} \right),_{r} 
+ \frac{2}{r} \left(\frac{ rh,_{r}}{h}\right),_{r} \nb\\
&& - \frac{ h(hG,_{tt} + 5 G,_{t}h,_{t})}{2G} 
 - \frac{(h^{2}h,_{t}),_{t}}{h}, \\
\lb{eq2.8}
R_{22} &=& \sin^{- 2}\theta R_{33} = \frac{r^{2}}{2} \left\{ 
\frac{(h^{3}G,_{r}),_{r}}{h^{3}G}  + \frac{(h^{6}G^{4}),_{r}}
{rh^{6}G^{4}}
+ \frac{2h,_{rr}}{h} \right.\nb\\
&& - \left. \frac{(h^{5}G,_{t}),_{t}}{ h^{3} G} 
 - \frac{2 (h^{2}h,_{t}),_{t}}{h}\right\}.
\eqn
To solve the Einstein field equations, we need to specify
the matter fields. In the following we shall consider two particular
cases, one is  a massless scalar field, and the other is 
 a radiation fluid. 

\vspace{1.cm}

\centerline{\bf A. Exact solutions of massless scalar field}

The Einstein field
equations for the massless scalar field can be written as
\bq
\lb{eq2.9}
R_{\mu\nu} = - 8\pi \phi,_{\mu} \phi,_{\nu}.
\eq
Because of the spherical symmetry, without loss of generality, we
assume that $\phi = \phi(t, r)$. Then, the above equation immediately
yield $R_{22} = 0.$ In view of Eq.(\ref{eq2.8}), this can be written as
\bq
\lb{eq2.10}
\frac{(h^{3}G,_{r}),_{r}}{h^{3}G}  + \frac{(h^{6}G^{4}),_{r}}
{rh^{6}G^{4}} + \frac{2h,_{rr}}{h} 
 = \frac{(h^{5}G,_{t}),_{t}}{ h^{3} G} 
+ \frac{2 (h^{2}h,_{t}),_{t}}{h}.
\eq
To solve this equation, let us consider Types $A$
and $B$ solutions separately.

{\bf Type A Solutions}. It can be shown that for Type $A$
solutions, Eq.(\ref{eq2.10}) has the general solution,
\bq 
\lb{eq2.11}
G(t, r) = c_{1} - \frac{f,_{u}(u) - g,_{v}(v)}{r^{2}} + \frac{f(u)
+ g(v)}{r^{3}},
\eq
where $c_{1}$ is an arbitrary constant, $f(u)$ and $g(v)$ are
arbitrary functions of their arguments, with $u \equiv t + r$ and $v
\equiv t - r$. 
It should be noted that when other components of the Einstein field
equations (\ref{eq2.5}) $-$ (\ref{eq2.7})  are considered, $f(u)$ and
$g(v)$ are not really arbitrary. They have to satisfy the
integrability condition for the massless scalar field, which can be
written as
\bq 
\lb{eq2.12}
R_{00} R_{11} - R_{01}^{2} = 0.
\eq
Once this condition is satisfied, the massless scalar field can be
obtained by integrating the other two independent field equations
(\ref{eq2.5}) and (\ref{eq2.7}), which can be written in the form
\bq 
\lb{eq2.13}
\phi,_{t} ^{2} = - \frac{1}{8\pi}R_{00}, \;\;\;\;\;\;\;\;\;  
\phi,_{r} ^{2} = - \frac{1}{8\pi}R_{11}.
\eq
To solve Eqs.(\ref{eq2.12}) and (\ref{eq2.13}) for the general solution
of $G$ given by Eq.(\ref{eq2.11}) turns out to be complicate.
Therefore, in the following we consider some particular solutions, which
are sufficient for our present purpose.

{\bf Case} $\alpha)$ If we choose the two functions $f$ and $g$ as
$$
f(u) = \frac{c_{2}u^{4}}{16}, \;\;\;\; \;\;\;\;  
g(v) = - \frac{c_{2}v^{4}}{16},
$$
where $c_{2}$ is an arbitrary constant, we have
\bq 
\lb{eq2.14}
G(t, r) =  c_{1} - c_{2} t.
\eq
For such a choice, one can show that Eq.(\ref{eq2.12}) is automatically
satisfied, while the integration of Eq.(\ref{eq2.13}) yields
\bq 
\lb{eq2.15}
\phi(t, r) = \pm \left(\frac{3}{16\pi}\right)^{1/2}
\ln \left| c_{1} - c_{2}t \right| + \phi_{0}, 
\eq
where $\phi_{0}$ is an integration constant.  Note that this solution
was also obtained recently in \cite{HMN1994}.  
 
{\bf Type B Solutions}. For Type B solutions, it can be shown that
Eq.(\ref{eq2.10}) has solutions only when $f_{1}(t)$ is a constant, and
they are given by
\bqn
\lb{eq2.17}
G(t, r) &=& \sum_{n=0}^{\infty}
  r^{- 2} \cosh\left[(k_{n} - 16f_{1})^{1/2}(t + t_{0})\right] 
(f_{1} + r^{2})^{3/2}\nb\\
&& \exp\left\{\frac{1}{2}N(r)\right\}
\left(a_{n} + b_{n}\int{e^{N(r)} dr}\right) , 
\eqn
where
 \bqn
\lb{eq2.18}
 N(r) &\equiv& 2\ln r - \ln(f_{1} + r^{2}) - 
2\ln\left[(f_{1} - r^{2}) - (k_{n} - 16 f_{1})^{1/2} r\right]  \nb\\
  && - 2 \left(\frac{k_{n} - 16f_{1}}{f_{1}}\right)^{1/2} Arctan 
 \left(\frac{r}{\sqrt{f_{1}}}\right),
\eqn
and $t_{0}, k_{n}, a_{n}$ and $b_{n}$ are arbitrary constants, subject
to Eq.(\ref{eq2.12}). 

{\bf Case} $\beta )$ One particular solution of Eq.(\ref{eq2.17})
 is given by
\bq
\lb{eq2.19}
G(t, r) = A \cosh(4\alpha t) - B \sinh(4\alpha t ), 
\eq
where $\alpha \equiv \sqrt{- f_{1}}\; (f_{1} < 0)$, $A$
and $B$ are two arbitrary constants.  For such a choice it can be
shown that the integrability condition (\ref{eq2.12}) is satisfied, and
Eq.(\ref{eq2.13}) has the solution
\bq
\lb{eq2.20}
\phi(t, r) = \pm \left(\frac{3}{16\pi}\right)^{1/2}
\ln\left|\frac{(A - B) e^{2\alpha t} - (B^{2} - A^{2})^{1/2}
e ^{- 2\alpha t}}{ (A - B) e^{2\alpha t} + (B^{2} - A^{2})^{1/2}
e ^{- 2\alpha t}}\right| + \phi_{0}, 
\eq 
where $\phi_{0}$ is another integration constant. Since $\phi$ is real,
we require $B^{2} - A^{2} \ge 0$.

\vspace{1.2cm}

\centerline {\bf B. Exact solutions of radiation fluid} 

For a perfect fluid, the energy-momentum tensor takes the form
$T_{\mu\nu} = (\rho + p)u_{\mu}u_{\nu} - p g_{\mu\nu}, $ where $u_{\mu}$
denotes the four velocity of the fluid.  In the present case, we can
assume that it has only two non-vanishing components, $u_{\mu} =
\{u_{0}, u_{1}, 0, 0\}.$ Then, one can show that only four of the ten
Einstein field equations are independent, and can be written in the form
\cite{Walker1935}
\bqn
\lb{eq2.21}
(R^{0}_{0} &-& R^{2}_{2})(R^{1}_{1} - R^{2}_{2}) - R^{0}_{1}R^{1}_{0} 
= 0, \\
\lb{eq2.22}
\rho &=& - \frac{1}{16\pi}(R^{0}_{0} + R^{1}_{1} - 4 R^{2}_{2}), \\
\lb{eq2.23}
p &=& - \frac{1}{16\pi}(R^{0}_{0} + R^{1}_{1}), \\
\lb{eq2.24}
u_{0}^{2} &=&  \frac{g_{00}(R^{0}_{0} - R^{2}_{2})}
{R^{0}_{0} + R^{1}_{1} - 2 R^{2}_{2}}.
\eqn
For the radiation fluid, we have $p = \rho/3$. Then, Eqs.(\ref{eq2.22})
and (\ref{eq2.23}) give
\bq
\lb{eq2.25}
 R^{0}_{0} + R^{1}_{1} + 2 R^{2}_{2} = 0.
\eq
In the following, we shall first solve Eq.(\ref{eq2.25}) to get the
general form of $G$, and then consider  the constraint equation
(\ref{eq2.21}). Once this is done, we shall use Eqs.(\ref{eq2.22}) and
(\ref{eq2.24}) to get the energy density of the fluid and its four
velocity. It can be seen that such obtained $\rho$ is not always
positive. Therefore, to have physical meaningful solutions, we shall
further impose the condition $\rho \ge 0$.

Before proceeding further, we would like to mention that all the
conformally flat perfect fluid solutions are known \cite{Kramer1980}.
However, here we shall re-derive them in a different system of
coordinates (\ref{eq2.1}) for the convenience of the study of their
gravitational collapse.

{\bf Type A Solutions}. When $h(t, r) = 1$, Eq.(\ref{eq2.25}) takes
the form
\bq
\lb{eq2.26}
\left( \frac{r^{2} G,_{r}}{G^{1/2}} \right),_{r}  =  
\left( \frac{r^2 G,_{t}} {G^{1/2}}\right),_{t} ,
\eq
which has the general solution,
\bq
\lb{eq2.27}
G(t, r) =  \left[\frac{f(u) + g(v)}{r}\right]^{2},
\eq
where $f(u)$ and $g(v)$ are two arbitrary functions, subjected to
Eq.(\ref{eq2.21}) and the condition $\rho \ge 0$.  

{\bf Case $\gamma )$} If we choose $f(u)$ and $g(v)$ as
$$
f(u) = \frac{c_{2}}{4}\left(u - \frac{c_{1}}{c_{2}}\right)^{2},   
\;\;\;\;\;\;
g(v) = - \frac{c_{2}}{4}\left(v - \frac{c_{1}}{c_{2}}\right)^{2},
$$
where $c_{1}$ and $c_{2}$ are  arbitrary constants, we find
\bq
\lb{eq2.28}
G(t, r) = (c_{1} - c_{2} t)^{2}.
\eq
For such a choice, one can easily show that Eq.(\ref{eq2.21}) is
satisfied, while Eqs.(\ref{eq2.22}) and (\ref{eq2.24}) give
\bq
\lb{eq2.29}
\rho = 3p = \frac{c_{2}^{4}}{8\pi (c_{1} - c_{2} t)^{4}}
, \;\;\;\;\;\;
u_{0} =  \frac{1}{c_{1} - c_{2} t}.
\eq
The above solution belongs to the general Friedmann-Robertson-Walker
solutions.

{\bf Type B Solutions}. When $ h(t, r) = [f_{1}(t) +
r^{2}]^{- 1}$, Eq.(\ref{eq2.25}) takes the form
\bqn
\lb{eq2.32}
G,_{t}^{2} &-& 2GG,_{tt} + \frac{6 Gf_{1},_{t} G,_{t} }{f_{1} + r^{2}}
- \frac{12G^{2}f_{1},_{t}^{2}}{(f_{1}+r^{2})^{2}} +  
\frac{4G^{2}f_{1},_{tt}}{f_{1}+r^{2}} - 16f_{1}G^{2} = \nb\\
&-& (f_{1} + r^{2})^{2}\left[2GG,_{rr} - G,_{r}^{2} + 
\frac{4f_{1}G G,_{r}}{r(f_{1} + r^{2})} \right].
\eqn
As in the case of the massless scalar field, the above equation has
solutions only when $f_{1}(t)$ is a constant, and the corresponding
solutions are given by
\bqn
\lb{eq2.33}
G(t, r) &=& \sum_{n=0}^{\infty}\cosh^{2}[k_{n}(t + t_{0})]
\frac{(f_{1} + {r}^{2})}{r^{2}}\exp[N(r)]\nb\\
&&\left(a_{n} + b_{n}\int{\exp[- N(r)] dr}\right)^{2},
\eqn
where
\bq
\lb{eq2.34}
N(r) \equiv \ln[2 (f_{1} + r^{2})] + \left(\frac{{4k_{n}}}{{f_{1}}} 
\right)^{1/2}Arctan\left(\frac{r}{\sqrt{f_{1}}}\right),
\eq
and $k_{n}, a_{n}$ and $b_{n}$ are arbitrary constants.

{\bf Case $\delta$)} A particular case of Eq.(\ref{eq2.33}) 
that satisfies Eq.(\ref{eq2.21}) is
\bq
\lb{eq2.35}
G(t, r) = [A \cosh(2\alpha t) - B\sinh(2\alpha t)]^{2},
\eq
where $A$ and $B$ are two constants, and $\alpha$ is defined as that in 
Case $\gamma)$. 
The corresponding physical quantities are 
\bqn
\lb{eq2.36}
\rho = 3p &=& \frac{3\alpha^{2}(B^{2} - A^{2})}
{2\pi [A\cosh(2\alpha t) - B \sinh(2\alpha t)]^{4}},\nb\\
u_{0} &=& \frac{1}{A\cosh(2\alpha t) - B \sinh(2\alpha t)}. 
\eqn

\section*{III. PHYSICAL INTERPRETATION OF THE EXACT SOLUTIONS}

\renewcommand{\theequation}{3.\arabic{equation}}
\setcounter{equation}{0}

To study the solutions given in the last section, let us first 
introduce the mass function $m(t, r)$ via the relation \cite{PI1990}
\bq
\lb{eq3.1}
1 - \frac{2 m(t, r)}{R} = - R,_{\alpha}R,_{\beta}g^{\alpha\beta},
\eq
where $R$ is the physical radius of the two-sphere $t, r = Const.$,
and is defined as that in Eq.(\ref{eq2.3}). On the apparent
horizon
\bq
\lb{eq3.2}
R,_{\alpha}R,_{\beta}g^{\alpha\beta} = 0,
\eq
the mass function reads
\bq
\lb{eq3.3}
M_{AH}(t, r) = \frac{R_{AH}}{2},
\eq
where $R_{AH}$ is a solution of Eq.(\ref{eq3.2}).

\vspace{1.cm}

\centerline{\bf A. Massless scalar field}

{\bf Case $\alpha$)}: In this case, the solutions are given by
Eqs.(\ref{eq2.14}) and (\ref{eq2.15}) with $h = 1$. According to the
values of the constant $c_{1}$, the solutions can be further divided into
two sub-cases, $c_{1} > 0$ and $c_{1} < 0$. However, it is easy to show
that with the replacement $t$ by $- t$, we can get one from the other. Thus,
without loss of generality, we need only consider the case $c_{1} > 0$.
Introducing a quantity $P \equiv c_{2}/c_{1}$, the metric can be written
as $ds^{2} = c_{1}(1 - Pt) ds^{2}_{M}$, where $ds^{2}_{M}$ denotes the
Minkowski metric. From this expression we can see that the amplitude of
$c_{1}$ does not play any significant role, hence in the following we shall
set $c_{1} = 1$. Then, the metric takes the form
\bq
\lb{eq3.4}
ds^{2} = (1 - Pt)[dt^{2} - dr^{2} - r^{2} d^{2}\Omega].
\eq
From Eqs.(\ref{eq3.1}) and (\ref{eq3.2}) we find that the corresponding 
mass function is given by
\bq
\lb{eq3.5}
m(t, r) = \frac{P^{2}r^{3}}{8(1 - Pt)^{3/2}},  
\eq
and that the apparent horizon is located on the hypersurface
\bq
\lb{eq3.6}
r^{2}_{AH} = \frac{4(1 - Pt)^{2}}{P^{2}}. 
\eq
On the other hand, the corresponding Kretschmann scalar is given by
$$
{\cal{R}} \equiv R_{\alpha\beta\gamma\delta}
R^{\alpha\beta\gamma\delta} = \frac{45P^{4}}{8(1 - Pt)^{6}},
$$
which shows that the spacetime is singular on the hypersurface $t = P^{-
1}$.  The nature of the singularity depends on the signature of the
parameter $P$.  In fact, when $P > 0$ the singularity is hidden behind
 the apparent horizon given by Eq.(\ref{eq3.6}), and the solution
represents the formation of black holes, due to the collapse of the
massless scalar field. The corresponding Penrose diagram is given by
Fig.1(a). From Eqs.(\ref{eq3.5}) and (\ref{eq3.6}) we can see that the
 total mass of the black hole is infinitely large. To get a black hole
with finite mass, we can cut the spacetime along a hypersurface and then
join it to an asymptotically flat exterior \cite{WO1996}. Equation
(\ref{eq2.15}) shows that the coordinates are comoving with the massless
 scalar field. Thus, we can cut the spacetime along the hypersurface $r
= r_{0}$, where $r_{0}$ is a constant, and then join the part $r \le
r_{0}$ to an asymptotically-flat exterior. Once this is done, the total
mass of the scalar field that falls inside the black hole is given by
Eq.(\ref{eq3.5}) at the point where the hypersurface $r = r_{0}$
intersects with the apparent horizon (\ref{eq3.6}), which is
\bq
\lb{eq3.7}
M_{BH} = \left(\frac{r^{3}_{0}}{8}\right)^{1/2} P^{1/2}. 
\eq

When $P = 0$, the massless scalar field $\phi$ becomes a constant, and
the corresponding metric (\ref{eq3.4}) becomes that of Minkowski.

When $P < 0$, the solutions are actually the time inverse of the ones with
$P > 0$. Thus, they represent  white holes, and the corresponding Penrose
diagram is given by Fig.1(b).

The above analysis shows that, although this class of solutions does not
exhibits critical phenomenon in the sense of Choptuik \cite{Ch1993}, the
mass of black holes does take a scaling form (\ref{eq3.7}) with the
exponent $\gamma$ being $0.5$, which is exactly the same as that given in the
collapse of a massless scalar wave packet with CSS
\cite{Brady1994,WO1996}. This result is a little bit surprising.
However, a closer exam of the solutions shows that they also have CSS.
In fact, Eq.(\ref{eq2.3}) has the solution
$$
\xi^{0} = - \frac{2(1 - Pt)}{3P},\;\;\;\;\; 
\xi^{1} = \frac{2 r}{3}.
$$
Moreover, introducing two new coordinates
$$
\bar{t} = \frac{2(1 - Pt)^{3/2}}{3P},\;\;\;\; \bar{r} = r^{3/2},
$$
the metric (\ref{eq3.4}) can be written as
$$
ds^{2} = d\bar{t}^{2} - \left(\frac{4Px}{9}\right)^{2/3}
d\bar{r}^{2} - \left(\frac{2Px}{3}\right)^{2/3}
\bar{r}^{2} d^{2} \Omega,
$$
where $x \equiv \bar{t}/\bar{r}$ is the self-similar variable. The above
expression takes the exact form for the solutions with CSS \cite{CT1971}.

Solutions with CSS were extensively studied by Brady \cite{Brady1995},
and all of them were divided into two classes. One can show that the
above solutions belong to Brady's second class. To show this, let us write
Eq.(\ref{eq2.15}) as
$$
\phi = \pm \kappa( \ln|x| - \ln\bar{r}) + \bar{\phi}_{0},
$$
where $\bar{\phi}_{0}$ is another constant, and $\kappa \equiv
(12\pi)^{- 1/2}$.  Thus, we have $4\pi\kappa^{2} = 1/3 < 1$, which falls
into Class II solutions of Brady \cite{Brady1995}. It should be noted
that Class II solutions were further divided into three sub-classes: i)
$4\pi \kappa_{c}^{2} < 4\pi\kappa^{2} < 1$; ii) $0 < 4\pi\kappa^{2} \le
4\pi\kappa_{c}^{2}$; and iii) $\kappa = 0$, where $\kappa_{c}$ is an
undetermined constant. In case i), it was conjectured that the collapse
always forms black holes, while in cases ii) and iii) critical phenomena
exist. The only difference between the last two cases is that in
case ii) the critical solution separates black holes from naked
singularities, while in case iii) it separates black holes from those
solutions that represent the dispersion of the collapse \cite{Brady1994}.
Clearly, our above solutions belong to case i).

{\bf Case $\beta)$} In this case the solutions are given by
Eqs.(\ref{eq2.19}) and (\ref{eq2.20}) with $h = (r^{2} -
\alpha^{2})^{-1}$. Since $B^{2} - A^{2} \ge 0$, we can introduce a
constant $t_{0}$ such that $\sinh(4\alpha t_{0}) = A/\sqrt{B^{2} -
A^{2}}$. Then, the metric coefficient $G(t, r)$ can be written as $G(t,
r) = (B^{2} - A^{2})^{1/2}\sinh[4\alpha(t_{0} - \epsilon t)]$, where
$\epsilon = sign(B)$. Clearly, the factor $(B^{2} - A^{2})^{1/2}$ 
does not play any significant role to
the properties of the solutions. Without loss of generality, we shall
set it equal to one.  On the other hand, introducing a new radial
coordinate $\bar{r}$ by 
$$
\bar{r} = - \int{h(t, r) dr} = \frac{1}{2\alpha}\ln
\left|\frac{\alpha + r} {\alpha - r}\right|,
$$
the corresponding metric becomes
\bq
\lb{eq3.11}
ds^{2} = \sinh[4\alpha(t_{0} - \epsilon t)]\left\{
dt^{2} - d\bar{r}^{2} - \frac{\sinh^{2}(2\alpha 
\bar{r})}{4\alpha^{2}}
d^{2}\Omega\right\},
\eq
and Eq.(\ref{eq2.20}) reads
\bq
\lb{eq3.12}
\phi(t, \bar{r}) = \pm \left(\frac{3}{16\pi}\right)^{1/2}
\ln\left|\tanh[2\alpha(t_{0} - \epsilon t)]\right| + \bar{\phi}_{0},
\eq
where $\bar{\phi}_{0}$ is another constant. To have a correct signature
for the metric, we require $t_{0} - \epsilon t \ge 0$.

The physical relevant quantities now are given by
\bqn
\lb{eq3.13}
m(t, \bar{r}) &=& \frac{\sinh^{3}(2\alpha \bar{r})}
{4\alpha\sinh^{3/2}[4\alpha(t_{0} - \epsilon t)]}, \nb\\
{\cal{R}} = R_{\alpha\beta\gamma\delta}
R^{\alpha\beta\gamma\delta} &=&
\frac{1440\alpha^{4}}{\sinh^{4}[4\alpha(t_{0} - \epsilon t)]},
\eqn
while the apparent horizon is given by
\bq
\lb{eq3.14}
\bar{r}_{AH} = 2 (t_{0} - \epsilon t), \;\; (\epsilon = \pm 1).
\eq
Equation (\ref{eq3.13}) shows that the solutions are singular on the
hypersurface $t_{0} - \epsilon t = 0$. When $\epsilon = + 1$, the
singularity is hidden behind the apparent horizon, and the solutions
represent the formation of black holes. The corresponding Penrose
diagram is similar to that of Fig.1(a). When $\epsilon = - 1$, the
apparent horizon is behind the singularity, and the solutions
represent white holes. The corresponding Penrose diagram is 
that of Fig.1(b). Thus, only the solutions with $\epsilon  = + 1$
represent the gravitational collapse of the massless scalar field.
Eqs.(\ref{eq3.13}) and (\ref{eq3.14}) show that in this case the total
mass of the black hole is also infinitely large. To obtain a black hole with
finite mass, we can cut the spacetime and
then join it to an asymptotically-flat exterior. Equation (\ref{eq3.12})
shows that the scalar field depends only on $t$. That is, the coordinate
system $\{t, \bar{r}, \theta, \varphi\}$ is comoving. Therefore,  
we can cut the spacetime along the hypersurface $\bar{r} =
\bar{r}_{0}$, where $\bar{r}_{0}$ is a constant.  Once this is done, the
total mass that the scalar wave packet falls inside the black hole
should be given by Eqs.(\ref{eq3.13}) and (\ref{eq3.14}) 
with $t_{0} - t = \bar{r}_{0}/2$,
namely,
\bq
\lb{eq3.15}
M_{BH} = \frac{\sinh^{3/2}(2\alpha\bar{r}_{0})}{4\alpha},
\eq
which, unlike the last case, is finite and non-zero for any given
scalar wave packet. Therefore, this model represents the formation of
black holes, which turns on always at finite masses.

It can be shown that this class of solutions does not have either
CSS or DSS.

\vspace{1.cm}

\centerline{\bf B. Radiation fluid}  

{\bf Case $\gamma)$} In this case the solutions are given by
Eqs.(\ref{eq2.28}) and (\ref{eq2.29}) with $h = 1$. Similar to Case
$\alpha$, now only the parameter $c_{2}$ is essential. Thus, without
loss of generality, we set $c_{1} = 1$ and $P = c_{2}$. Then, the
physically relevant quantities are 
\bqn
\lb{eq3.16}
m(t, r) &=& \frac{P^{2}r^{3}}{2(1 - Pt)},  \nb\\
{\cal{R}} \equiv R_{\alpha\beta\gamma\delta}
R^{\alpha\beta\gamma\delta} &=& \frac{36P^{4}}{(1 - Pt)^{8}},
\eqn
while the apparent horizons now are located on
\bq
\lb{eq3.17}
r_{AH} = \frac{|1 - Pt|}{|P|}.
\eq
Equation (\ref{eq3.16}) shows that the spacetime is singular on the
hypersurface $1 -  Pt = 0$. The nature of the singularity depends on the
signature of $P$. In fact, when $P > 0$, it is space-like and hidden
behind the apparent horizon. The corresponding solutions represent the
formation of black holes. The Penrose diagram is given by Fig.2(a),
which is quite similar to Fig.1(a), except that now the apparent
horizon is null.  When $P = 0$ the metric reduces to Minkowski. When
$P < 0$, the singularity preceeds the apparent horizon, and the
corresponding solutions represent white holes [cf. Fig.2(b)].

The mass function (\ref{eq3.16}) on the apparent horizon (\ref{eq3.17})
takes the form
\bq
\lb{eq3.18}
M_{AH} = \frac{P r_{AH}^{2}}{2},
\eq
which diverges as $r_{AH} \rightarrow + \infty$. That is, the total
masses of black holes are infinite. To obtain black holes with finite
masses, following the previous cases, we can cut the spacetime along the
hypersurface $ r = Const.$, say, $r_{0}$, since now the radiation fluid
is comoving, too [cf. Eq.(\ref{eq2.29})]. Then, we can see that the
total mass that the fluid falls inside the black hole is given by
Eq.(\ref{eq3.18}) with $r_{AH} = r_{0}$,
\bq
\lb{eq3.19}
M_{BH} = \frac{r_{0}^{2}}{2} P.
\eq
That is, in this case $M_{BH}$ also takes a scaling form but with the
exponent $\gamma$ being equal to $1$. This is different from the value
$\gamma \approx 0.36$ found in \cite{AE1994,Koike1995}. 

Note that the
solutions studied here and the ones studied in \cite{AE1994,Koike1995}
all have CSS. As a matter of fact, Eq.(\ref{eq2.3}) has the solution
$$
\xi^{0} = - \frac{1 - Pt}{2P}, \;\;\;
\xi^{1} = \frac{r}{2}.
$$ 
Therefore, the difference between the values of the exponent $\gamma$
are not due to the different self-similarities, as that in Case
$\alpha$. We believe that this is due to the regular condition at the
center. In \cite{Koike1995} it was shown that if the solutions are
analytic and have a regular center, the mass of black holes with CSS
must take a scaling form with $\gamma \approx 0.36$. Thus, we
conjecture that if we give up the analytic condition, replacing, for
example, by the condition that the metric is $c^{1}$, as did in
\cite{Brady1994}, one should find solutions that represent critical
collapse with a scaling form of mass and $\gamma = 1$. Since in the
present case the solutions do not represent the critical collapse, we
can not verify this point here.

{\bf Case $\delta)$} In this case the solutions are given by
Eqs.(\ref{eq2.35}) and (\ref{eq2.36}) with $h(t, r) = (r^{2} -
\alpha^{2})^{- 1}$. Similar to Case $\beta$, we can introduce a
constant $t_{0}$ by $\sinh(2\alpha t_{0}) = A/(B^{2} - A^{2})^{1/2}$,
and write the metric in the form
\bq
\lb{eq3.22}
ds^{2} = \sinh^{2}[2\alpha(t_{0} - t)]\left\{ dt^{2} - d\bar{r}^{2}
- \frac{\sinh^{2}(2\alpha\bar{r})}{4\alpha^{2}}\;d^{2}\Omega\right\},
\eq
where $\bar{r}$ is defined as that in Case $\beta$. Then, we find that
\bqn
\lb{eq3.23}
m(t, \bar{r}) &=& \frac{\sinh^{3}(2\alpha \bar{r})}
{4\alpha\sinh[2\alpha(t_{0} - \epsilon t)]}, \nb\\
{\cal{R}} = R_{\alpha\beta\gamma\delta}
R^{\alpha\beta\gamma\delta} &=&
\frac{576\alpha^{4}}{\sinh^{8}[2\alpha(t_{0} - \epsilon t)]},
\eqn
and
\bq
\lb{eq3.24}
\bar{r}_{AH} = t_{0} - \epsilon t,
\eq
where $\epsilon \equiv sign(B)$. The above expressions show that when
$\epsilon = + 1$, the solutions represent the formation of black holes,
and the corresponding Penrose diagram is that of Fig.2(a). When
$\epsilon = - 1$, they represent white holes.  The corresponding Penrose
diagram is that of Fig.2(b).

The masses of black holes are infinite, as we can see from the above
expressions.  But, since the coordinates are comoving with the fluid [cf.
Eq.(\ref{eq2.36})], we can cut the spacetime along the hypersurface
$\bar{r} = \bar{r}_{0}$, where $\bar{r}_{0}$ is a constant, and then
join the part $\bar{r} \le \bar{r}_{0}$ with an asymptotically-flat
exterior. Once this is done, the mass that the fluid falls inside the
black hole is
\bq
\lb{eq3.25}
M_{BH} = \frac{\sinh^{2}(2\alpha \bar{r}_{0})}{4\alpha},
\eq
which is finite and non-zero for any given collapsing shell of
radiation fluid.  Therefore, this case also represents the formation
of black holes, which turns on with finite masses.

Using Eqs.(\ref{eq2.3}) and (\ref{eq2.4}), one can show that the
solutions in this case have neither CSS nor DSS.

\section*{IV. CONCLUDING REMARKS} 

In this paper, we have presented several classes of conformally flat and
spherically symmetric exact solutions to the Einstein field equations,
coupled with either a massless scalar field or a radiation fluid.  Some
of these solutions represent the formation of black holes, due to the
gravitational collapse of the matter fields. However, since the masses
of black holes are all infinite, we have discussed the possibility of
cutting the spacetime along a hypersurface, and then joining the
internal part with an asymptotically-flat exterior, so that the
resulting masses of black holes are finite. Once this is done, we have
shown that the masses of such formed black holes always take a scaling
form for spacetimes with CSS for both massless scalar field and
radiation fluid.  The corresponding exponent $\gamma$ is $0.5$ for the
massless scalar field, and $1$ for the radiation fluid. In
\cite{Koike1995} it was shown that the masses of black holes formed from
the critical collapse of radiation fluid with CSS {\em always} take a
scaling form but with $\gamma \approx 0.36$. This seems to contradict
with the results obtained here. However, it should be pointed out that
the results obtained in \cite{Koike1995,Gundlach1995} are based
on the requirement that the solutions be analytic and have a regular
center.  When we give up one of the two conditions, we would expect that
the results would be different, in particular, we should be able to
construct solutions that represent critical collapse of a radiation
fluid with masses of black holes taking a scaling form and the exponent
being $1$. Of course, this is just a speculation, since our solutions
constructed here do not really represent critical collapse.  To clarify
this point, it would be very useful to consider solutions with less
requirements than those in \cite{Koike1995,Gundlach1995}.

On the other hand, the masses of black holes formed from the collapse
that has neither CSS nor DSS {\em always} turn on at finite values,
which supports our conjecture made in \cite{WO1996}. Thus, if
astrophysically interesting black holes are all with finite non-zero
mass, Nature seems to forbid solutions with CSS or DSS.

\section*{ACKONWLEDGMENT}

One of the authors (AW) would like to thank Henrique P. de Oliveira
for useful discussions. The financial assistance from CNPq is gratefully
 acknowledged.


\newpage

\section*{Figure Captions}

Fig.1 The Penrose diagram for Case $\alpha)$ defined in text: (a) for $P
> 0$ and (b) for $P < 0$. The spacetime singularities are represented by
dash lines.  The apparent horizons (AH) are space-like.

Fig.2 The Penrose diagram for Case $\gamma)$: (a) for $P > 0$ and (b)
for $P < 0$. The spacetime singularities, represented by dash lines,
are space-like, while the apparent horizons (AH) are null.


\begin{thebibliography}{100}

\bibitem{Penrose1969} R. Penrose, Riv. Nuovo Cimento 1, 252 (1969).

\bibitem{Chris1986} D. Christodoulou, Commun. Math. Phys. 105, 337
(1986); {\em ibid.} 106, 587 (1986); {\em ibid.} 109, 591 (1987); {\em
ibid.} 109, 613 (1987); Commun. Pure Appl. math. XLIV, 339 (1991).

\bibitem{GP1987} D.S. Goldwirth and T. Piran, Phys. Rev. D36, 3575
(1987).

\bibitem{Ch1993} M.W. Choptuik, Phys. Rev. Lett. 70, 9 (1993).

\bibitem{AE1994} A.M. Abrahams and C.R. Evans, Phys. Rev. Lett. 70,
2980 (1993); Phys. Rev. D49, 3998 (1994).

\bibitem{EC1994} C.R. Evans and J.S. Coleman, Phys. Rev. Lett.  72,
1782 (1994).

\bibitem{EH1995} E.W. Hirschmann and D.M. Eardley, Phys. Rev. D51, 4198
(1995); {\em ibid.} 52, 5850 (1995); D. Garfinkle, 
{\em ibid.} 51, 5558 (1995); R.S. Hamad\'e and J.M. Stewart, Class.
Quantum Grav. 13, 497 (1996).

\bibitem{Brady1994} P.R. Brady, Class. Quantum Grav. 11, 1255 (1994);
 Y. Oshiro, K. Nakamura, and A. Tomimatsu,
Prog. Theor. Phys. 91, 1265 (1994).

\bibitem{Koike1995} T. Koike, T. Hara, and S. Adachi,
Phys. Rev. Lett.  74, 5170 (1995); T. Koike, ``{\em Renormalization
group analysis of critical behavior in gravitational collapse}," Ph. D.
Thesis submitted, to Department of Physics, Tokyo Institute of Technology
(1996); T. Hara, T. Koike, and S. Adachi, ``{\em Renormalization group
and critical behavior in gravitational collapse}," gr-qc$/9607010$
(1996).

\bibitem{Gundlach1995} C. Gundlach, Phy. Rev. Lett. 75, 3214 (1995);
Phys. Rev. {\bf D55}, 695 (1997).



\bibitem{Maison1995} D. Maison, Phys. Lett. B366, 82 (1996).

\bibitem{KO1995} T. Koike and T. Mishima,  Phys. Rev. D51, 4045 (1995);
H.P. de  Oliveira and E.S. Cheb-Terrab, Class. Quantum Grav. 13, 425
(1996); H.P. de Oliveira, ``{\em Self-Similar Collapse in Brans-Dicke
Theory and Critical Behavior}," gr-qc$/$9605008; T. Chiba and J. Soda,
Prog. Theor. Phys. {\bf 96}, 567 (1996).

\bibitem{EHH1995}D.M. Eardley, E.W. Hirschmann, and J.H. Horne, Phys.
Rev. D52, {\bf R}5397 (1995);  E.W. Hirschmann and D.M. Eardley,
 ``{\em Criticality and Bifurcation in the Gravitational Collapse of a
Self-Coupled Scalar Field}," report, gr-qc/9511052 (1995).
 
\bibitem{WO1996} A.Z. Wang and H.P. de Oliveira, ``{\em Critical
Behavior of Collapsing Massless Scalar Wave Packets}", 
gr-qc/9608063 (1996); Phys. Rev. {\bf D55} (12), xxx (1997), in press.

\bibitem{Roberts1989} M.D. Roberts, Gen. Relativ. Grav. 21, 907 (1989).

\bibitem{LL1975} L.D. Landau and E.M. Lifshitz, {\em The Classical
Theory of Fields}, (Pergamon Press, New York, 1975), pp. 299-303.

\bibitem{BOS1989} W.B. Bonnor, A.K.G. de Oliveira, and N.O. Santos,
 Phys. Rep. 181, 269 (1989).

\bibitem{CT1971} M.E. Cahill and A.H. Taub, Commun. math. Phys. 21, 
1 (1971).  

\bibitem{HMN1994} V. Husain, E.A. Martinez, and D.  Nunez, Phys. Rev.
D50, 3783 (1994).

\bibitem{Walker1935} A.G. Walker, Quart. J. Math. 6, 81 (1935); W.B.
Bonnor and H. Knutsen, Inter. J. Theor. Phys. 32, 1061 (1993).

\bibitem{Kramer1980} D. Kramer, H. Stephani, E. Herlt, and M. MacCallum,
``{\em Exact Solutions of Einstein's Field Equations}", (Cambridge
University Press, Cambridge, 1980), p.371.

\bibitem{PI1990} E. Poisson and W. Israel, Phys. Rev. D41, 1796 (1990).

\bibitem{Brady1995} P.R. Brady, Phys. Rev. D51, 4168 (1995).
 
\end{thebibliography}
\end{document}